# Effectiveness of Large Language Models in Simulating Regional Psychological Structures: An Empirical Examination of Personality and Subjective Well-being

*Ke Luoma*[1], *Li Zengyi* [2], *Liao Jiangqun*[3], *Tong Song\*\**[4,5], *Peng Kaiping\*\**[1]

(1 Department of Psychological and Cognitive Science, Tsinghua University，Beijing，100084)

(2 Neoma Business School, France, 76130)

(3 Business School , Beijing Technology and Business University，Beijing，100048)

(4 Department of Psychology, Faculty of Arts and Sciences, Beijing Normal University, Zhuhai, 519087)

(5 Beijing Key Laboratory of Applied Experimental Psychology, Faculty of Psychology, Beijing Normal University, Beijing, 100875)

**Abstract**: The study aims to evaluate the capability of a large language model (DeepSeek) to simulate group-level psychological characteristics when conditioned on demographic features. Using a sample demographically matched to the China Family Panel Studies (CFPS, 2018; N = 2,943), we constructed AI-generated "virtual participants" and compared them with real participants to analyze regional differences in the Big Five personality traits and subjective well-being (SWB), as well as their associations. The results show that the simulated data broadly align with the real data in the regional distribution patterns of SWB and the Big Five, while exhibiting some idiosyncratic deviations in finer details; moreover, several personality dimensions predict SWB. These findings indicate the potential of LLMs such as DeepSeek to simulate regional psychological structures, while underscoring the need to attend to cultural sensitivity and fine-grained feature modeling. The study provides empirical support for assessing the effectiveness of LLMs in modeling population psychological characteristics.

**Keywords:** large language model　DeepSeek　Big Five personality　well-being　regional psychological structure　virtual participants

Corresponding authors: Tong Song (s.tong@bnu.edu.com); Peng Kaiping (pengkp@tsinghua.edu.cn)

Funding: National Key Research & Development Program of China (2016YFA0602500) and Tsinghua University Institute for Global Industry self-selected projects (2021-11-09-LXHT005-01; 2024-06-18-LXHT002).

DOI: 10.16719/j.cnki.1671-6981.20250412

# 1. Introduction

Large language models (LLMs) demonstrate strong abilities in language understanding and generation (Bubeck et al., 2023; Rathje et al., 2024), opening new opportunities for intelligent research on group psychological structures in social psychology (Ke et al., 2024). Prior work indicates that responses produced by GPT-4 can closely match human self-report scores on personality questionnaires (De Winter et al., 2024; Wang et al., 2025), implying a degree of "role-playing" capability whereby models can emulate agents with different personality profiles (Mei et al., 2024; Strachan et al., 2024). In computational social science, LLMs have also been explored as a means to generate "virtual participants," partially substituting human participants to reduce costs and simplify procedures (Bisbee et al., 2024; Dillion et al., 2023; Grossmann et al., 2023; Sarstedt et al., 2024).

Although prior work has examined the potential of LLMs to simulate group-level psychological structure, there is scant research on whether they can, at the group level, produce simulated data of sufficient quality—for example, reproducing real human structural differences and the regional distribution of psychological characteristics. Compared with reproducing individual-level personality, simulating group-level psychological structure is more challenging: it involves not only differences in the distributions of psychological variables, but also the stability and variability of these variables across different sociocultural ecologies. Cross-cultural psychology suggests psychological characteristics show non-random regional patterns influenced by social structures and cultural contexts (Talhelm et al., 2014). For example, Talhelm et al. (2014) compared university students from China's southern rice-growing regions with those from the northern wheat-growing regions and found that southerners tend to exhibit more holistic, interdependent thinking, whereas northerners are more analytical and independent. This result reflects the impact of regional cultural ecologies on psychological traits. Such differences may manifest in key variables such as personality scales and subjective well-being (Anglim et al., 2020).

The Big Five personality model provides a structured framework—openness, conscientiousness, extraversion, agreeableness, and neuroticism (McCrae & John, 1992; Paunonen & Ashton, 2001)—that relates to behaviors and health outcomes. Meta-analytic evidence links conscientiousness to lower health-risk behaviors and higher health-promoting behaviors (Bogg & Roberts, 2004). Well-being, meanwhile, is widely used to index quality of life and psychological adjustment (Schimmack et al., 2002, 2004). Personality traits are stably associated with well-being, and person-environment/behavioral "fit" also matters: when consumption aligns with personality, life satisfaction increases—even more than income predicts (Matz et al., 2016). Selecting personality and well-being as core indicators thus helps evaluate whether LLMs can capture both differential distributions and culturally sensitive structures in group-level simulations. Traditional personality research depends heavily on self-report scales, which are susceptible to social desirability and response style biases (Tourangeau & Yan, 2007). Cross-regional surveys incur high time and resource costs, and face challenges in sampling and geographic matching. Social psychology's reliance on self-report and limited direct behavioral/dynamic data (Baumeister et al., 2007) motivates exploring new measurement tools.

In recent years, applications of large language models (LLMs) in psychometrics have been increasingly explored. Preliminary evidence indicates that LLMs can generate responses that are highly correlated with participants' self-reports across scale dimensions (De Winter et al., 2024), suggesting a potential—under certain conditions—to simulate psychometric data and thus provide new theoretical support for developing virtual participants and modeling regional psychological structure (Hou et al., 2025).

However, alongside these advantages come potential risks and limitations. Studies have noted that LLM-generated virtual participants may tend to amplify existing social or cultural stereotypes (Argyle et al., 2023; Lucy & Bamman, 2021). This is especially salient in regional or cross-cultural research, because training corpora are largely sourced from public internet text, which often reinforces societal stereotypes about regional characteristics; consequently, model outputs may exaggerate or distort simulated regional psychological profiles (Demszky et al., 2023; Harding et al., 2024). Therefore, although LLMs create new opportunities for large-scale, cross-regional measurement of psychological structure, whether they can reproduce—at the level of group psychological structure—the regional distributions of personality and well-being remains to be validated through rigorous empirical testing, particularly regarding their ability to accurately simulate responses from individuals with complex group identities and to avoid flattening group characteristics (Wang et al., 2025).

Building on the above background and theoretical foundation, this study examines—across three levels of "personality traits, well-being, and regional culture"—the ability of large language models to simulate multi-level psychological structures (including the structure of psychological traits, regional psychological structure, and their interrelationships) and the biases therein. First, we test whether the regional distributional trends of well-being in LLM-generated virtual-participant data are consistent with those in real data. Second, we examine whether the personality data generated by the LLM can reproduce regional personality profiles that align with real data. Moreover, given the stable structural associations between the Big Five and well-being (Anglim et al., 2020; Steel et al., 2008), we further assess whether the relationships between each Big Five dimension and well-being in the LLM-generated data reflect the structural associations observed in human data.

This study combines large-scale survey data with AI-based generation methods to test whether the large language model DeepSeek can simulate regional differences in the Big Five personality traits and well-being among populations in China. In other words, we use DeepSeek to model the personality and well-being patterns characteristic of different Chinese regions. This approach provides new tools and research avenues for studying regional psychological structure and offers a feasible pathway for developing measurement tools for large-scale psychological surveys.

## 2. Methods

2.1 Real-world sample source

The real (human) sample was drawn from the China Family Panel Studies (CFPS). Launched in 2010 and fielded biennially, CFPS has completed five national waves to date. Questionnaire data can be requested free of charge via the project website[1].

In this study, we used personality and well-being data from the CFPS 2018 adult questionnaire, yielding 2,943 participants. The sample covers seven macro-regions—North China (Beijing, Tianjin, Hebei, Shanxi, Inner Mongolia), Northeast China (Liaoning, Jilin, Heilongjiang), East China (Shanghai, Jiangsu, Zhejiang, Anhui, Fujian, Jiangxi, Shandong), Central China (Henan, Hubei, Hunan), South China (Guangdong, Guangxi, Hainan), Southwest China (Chongqing, Sichuan, Guizhou, Yunnan, Tibet), and Northwest China (Shaanxi, Gansu, Qinghai, Ningxia, Xinjiang)—and was obtained via random sampling, with equal numbers of men and women, ages 18–65. Personality was assessed with the 15-item Chinese short Big Five scale (Hahn et al., 2012). Subjective well-being (SWB; hereafter SWB) was measured with a single self-report item: "How happy do you feel?" (0 = lowest happiness; 10 = highest).

2.2 AI virtual-sample generation strategy

The virtual sample in this study was generated with the DeepSeek-V3-0324 model and comprised 3,000 "participants," whose regional and demographic composition was matched to that of the human sample. First, in the chat window on the DeepSeek official website[2] we entered the following prompt (in Chinese): "You are now my assistant for a psychology experiment. Based on the GDP, cultural background, and demographic characteristics of China's 31 provinces/municipalities, generate information for xx participants by random sampling. For each participant, report: ID (xx–xx), province/municipality of residence, age (18–65), and gender (male or female). Ensure a balanced gender ratio. Please report all xx participants in full, without omission." Here, xx denotes the number of participants to be generated (e.g., generating 10 per batch implies xx = 10; the ID range would be 01–10), which controls the sample size and ID range for each simulation round. Next, using the participant information, we called the DeepSeek API (version V3-0324; temperature set to 0.7 to balance response diversity and consistency[3]) to have the model complete the Big Five and well-being questionnaires (Argyle et al., 2023), thereby producing the corresponding data. The questionnaire procedure complied with ethical guidelines; no real human respondents participated, and each questionnaire was generated independently.

The strategy of using an LLM to create simulated respondents for questionnaire administration has been explored in recent work (De Winter et al., 2024). Building on this, the present study expanded the sample size and measurement scope and compared the simulated data with data from real human participants.

Each "participant" received a system prompt containing personal information (ID, province, age, and gender) and responded from a first-person perspective. The questionnaires were presented in a fixed order with no item randomization. The prompt was as follows:

"Please role-play a character (ID, province/municipality, age, and gender). Drawing on your life experience, cultural background, and the prevailing social environment of your locale, imagine that you hold this identity and then answer the following questions about your personality and feelings as realistically as possible, from a first-person perspective. Maintain this identity throughout the entire questionnaire. You will answer two questionnaires. The first is the Big Five personality scale (items 1–15), which assesses your personality across five dimensions: Openness, Conscientiousness, Extraversion, Agreeableness, and Neuroticism. The second is the Life Satisfaction scale (items 16–17), which evaluates your overall satisfaction with life, including interpersonal relationships and well-being."

1. CFPS project website: http://www.isss.pku.edu.cn/cfps/

2. DeepSeek Chat: https://chat.deepseek.com/

3. The temperature setting ranges from 0–2 (see https://api-docs.deepseek.com/quick_start/parameter_settings) and controls the randomness of the model's text generation. Higher temperatures increase randomness and diversity, whereas lower temperatures yield more conservative and consistent responses.

2.3 Materials

The personality and subjective well-being (SWB) measures for the simulated "participants" were identical to those in the CFPS 2018 questionnaire. The short-form Big Five personality inventory comprised five dimensions—Openness, Conscientiousness, Extraversion, Agreeableness, and Neuroticism—with three items per dimension. For example, Extraversion items included "talkative," "cheerful, sociable," and

"reserved, conservative" (reverse-scored) (Wang et al., 2022). SWB was assessed with a single self-report item scored from 0 to 10 (0 = very unhappy; 10 = very happy).

Personality scores were computed as the mean score within each dimension; reverse-keyed items were scored as 5 minus the original response (Wu & Gu, 2020). Because the DeepSeek outputs matched the human response format, we applied the same preprocessing and corrections prior to analysis to ensure comparability. During data cleaning, simulated respondents with scores < 0—i.e., values outside the valid range of the scale generated by the model—were deemed invalid and excluded. In total, 57 such invalid simulated respondents were removed, leaving 2,943 simulated respondents whose scores fell within the valid range.

2.4 Data processing

Data analyses were conducted in Python 3.12 and R in three steps. First, independent-samples t tests were used to compare distributional differences between the real and simulated data. Second, to compare differences between the two datasets, we (a) computed descriptive statistics for the Big Five and subjective well-being (SWB) across regions for both the human and DeepSeek-simulated samples, and (b) performed one-way ANOVAs on personality dimensions and SWB across regions and compared the results. Third, multiple regression analyses were conducted to examine, in both datasets, the relationships between the Big Five dimensions and SWB.

## 3. Results

3.1 Comparison of human and simulated samples

The human sample comprised 2,943 individuals (male = 1,482; female = 1,461; age: M = 43.03 years, SD = 13.47). The DeepSeek-simulated sample also included 2,943 individuals (male = 1,482; female = 1,461; age: M = 42.92 years, SD = 11.82), and the two samples were closely matched demographically.

We used independent-samples t tests to compare the DeepSeek-simulated and human samples on the Big Five dimensions and subjective well-being (SWB). Results are reported in Table 1 and visualized in Figure 1. Although Conscientiousness differed significantly between groups (t = 2.67, p < .01), the corresponding effect size was trivial (Cohen's d = .07), suggesting that statistical significance was primarily driven by the large sample size and that the practical difference is limited. By contrast, the simulated sample scored significantly lower than the human sample on Extraversion (t = 43.98, p < .001, Cohen's d = 1.15) and Openness (t = 20.51, p < .001, Cohen's d = .53), with effect sizes in the medium-to-large range. For Agreeableness (t = 26.03, p < .001, Cohen's d = .68) and Neuroticism (t = 26.10, p < .001, Cohen's d = .68), the simulated sample scored significantly higher than the human sample, with similarly pronounced differences. For SWB (t = 22.79, p < .001, Cohen's d = .59), the simulated sample again scored significantly lower than the human sample, indicating a moderate discrepancy in the simulation of well-being.

To further evaluate the model's performance in simulating regional psychological structure, we plotted heatmaps of Cohen's d effect sizes across regions for the Big Five dimensions and subjective well-being (SWB) (see Figure 2). The results show that regional effect sizes for Conscientiousness are mostly near zero, indicating higher consistency on this dimension. In contrast, most other dimensions exhibit moderate to large deviations (Cohen's d > .5), suggesting room for improvement in modeling fine-grained regional characteristics.

**Table 1. Differences between DeepSeek-simulated and human samples on Big Five and well-being**

| Dimension | t | p | Cohen's d |
| --- | --- | --- | --- |
| C (Conscientiousness) | 2.67 | < .01 | .07 |
| E (Extraversion) | 43.98 | <.001 | 1.15 |
| A (Agreeableness) | 26.03 | <.001 | .68 |
| O (Openness) | 20.51 | <.001 | .53 |
| N (Neuroticism) | 26.10 | <.001 | .68 |
| SWB (Subjective Well-Being) | 22.79 | <.001 | .59 |

Note: the t-value is the test statistic for the independent-samples t test; Cohen's d represents the effect size (0.2 = small, 0.5 = medium, 0.8 = large) (Cohen, 1992).

Throughout the paper, we use the following abbreviations: N (Neuroticism), E (Extraversion), O (Openness), A (Agreeableness), C (Conscientiousness), and SWB (subjective well-being).

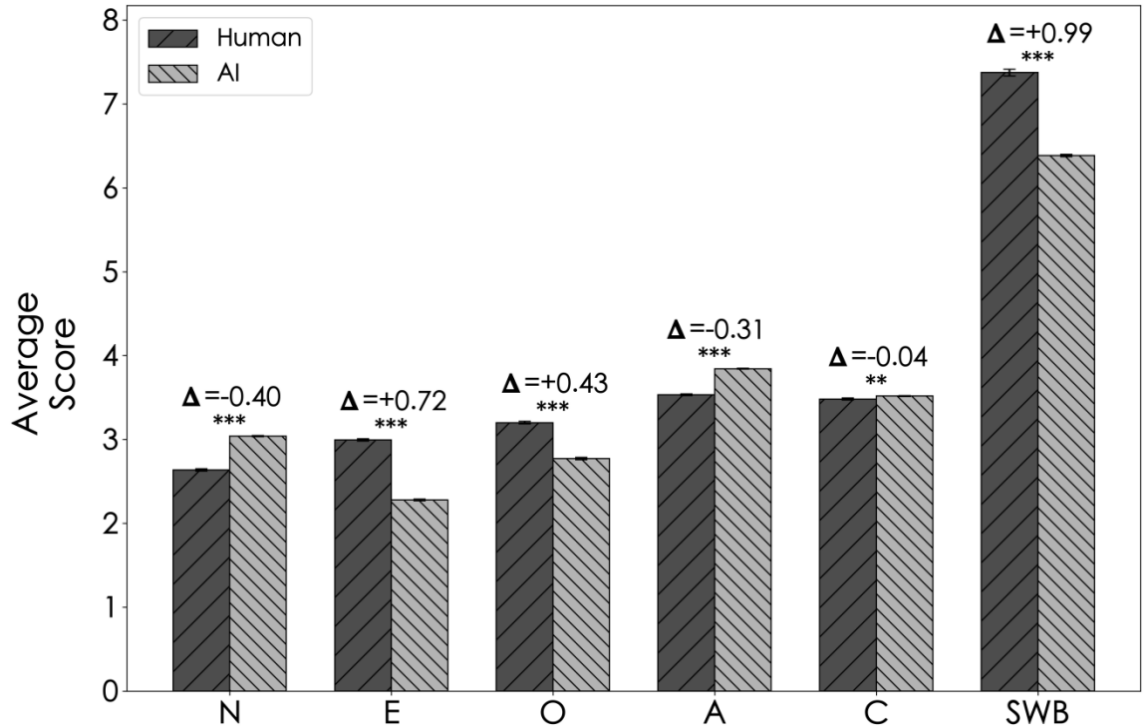

Figure 1. Big Five and well-being scores in AI vs. human samples and their differences.

Note. Δ = Human score − Simulated score (positive values indicate higher Human scores; negative values indicate higher Simulated scores); error bars indicate standard errors; * denotes p < .05, ** denotes p < .01, *** denotes p < .001; n.s. denotes not significant.

3.2 Regional differences

*3.2.1 Descriptive statistics: Personality and SWB*

We compared the human and AI-simulated samples on the Big Five and subjective well-being (SWB). The simulated sample scored significantly higher on Conscientiousness than the human sample, especially in East China (M_sim = 3.57 vs. M_human = 3.47). For Extraversion, the simulated sample was generally lower, with the largest gap in the Northwest (M_sim = 2.09 vs. M_human = 2.92). For Agreeableness, the simulated sample was generally higher, particularly in East (M_sim = 3.85 vs. M_human = 3.48) and South China (M_sim = 3.83 vs. M_human = 3.46). For Neuroticism, simulated scores were overall higher, most notably in the Northeast (M_sim = 3.12 vs. M_human = 2.60). For Openness, simulated scores were relatively lower, with the largest gap in the Northwest (M_sim = 2.52 vs. M_human = 3.29). For SWB, human scores were generally higher than simulated scores, with the largest gap in the Northeast (M_human = 7.71 vs. M_sim =

6.24), exceeding one point. Table 2 summarizes the means (M) and standard deviations (SD) for the Big Five and SWB across China's seven macro-regions in the human and DeepSeek-simulated samples.

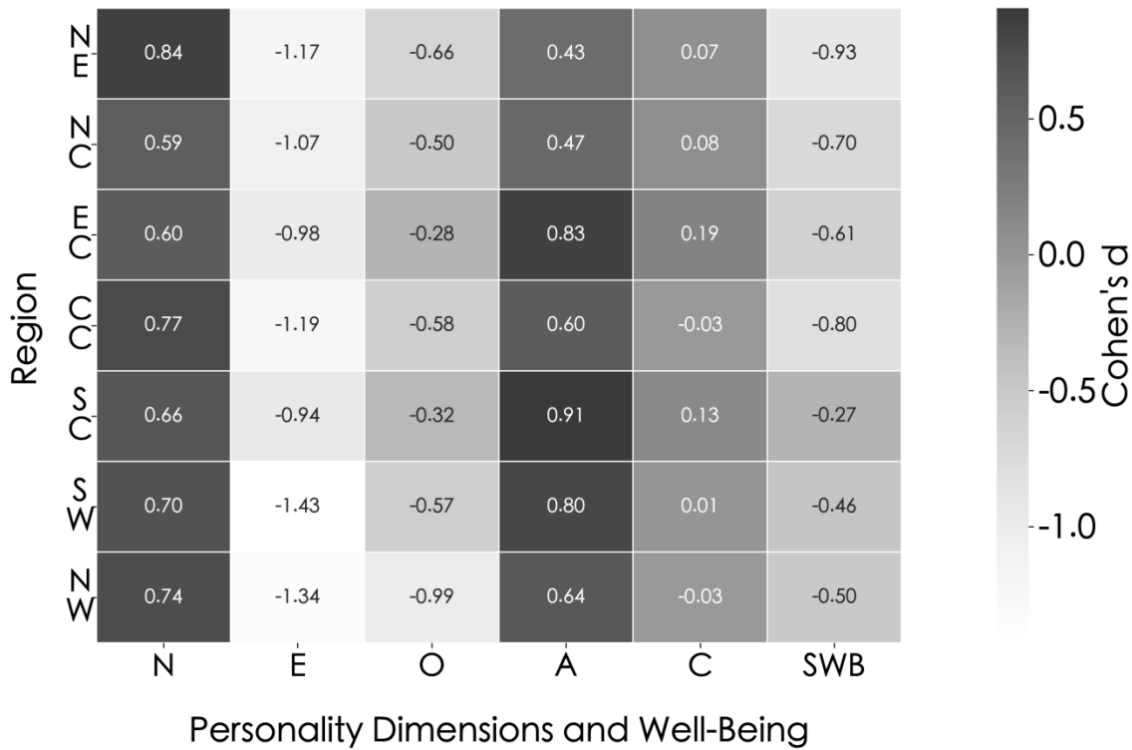

**Figure 2. Regional differences (Cohen's d) between simulated and human samples on Big Five and well-being.**

Note: Positive values indicate that the simulated sample scores higher than the Human sample; negative values indicate lower. Color intensity indicates the magnitude of the deviation; thresholds are in terms of Cohen's d: |d| < 0.2 denotes a small difference, 0.5–0.8 a medium difference, and |d| ≥ 0.8 a large effect (Cohen, 1992).

Throughout the paper, we use the following abbreviations: NE (Northeast); NC (North China); EC (East China); CC (Central China); SC (South China); SW (Southwest); NW (Northwest).

*3.2.2 One-Way ANOVAs for Personality and SWB*

To examine regional differences in personality dimensions and SWB, we conducted one-way ANOVAs separately for the human and DeepSeek-simulated samples. In the human sample, Extraversion and Agreeableness showed significant regional differences (Extraversion: $F(6, 2936) = 3.67$, $p < .01$, partial $\eta^2$

= .01; Agreeableness: $F(6, 2936) = 5.23$, $p < .001$, partial $\eta^2 = .01$), whereas the other three dimensions (Conscientiousness, Neuroticism, Openness) did not. SWB also differed significantly across regions ($F(6, 2936) = 6.24$, $p < .001$, partial $\eta^2 = .01$). Tukey post-hoc tests indicated that the Southwest scored significantly higher in Extraversion than the Northeast, East, and Northwest (mean differences = .20, .13, and .20, respectively; $ps < .05$); the Northeast scored significantly higher in Agreeableness than the East, South, and Southwest (mean differences = .14, .16, and .13; $ps < .05$). For SWB, the Northeast scored significantly higher than the South, Northwest, and Southwest (mean differences = .76, .58, and .57; $ps < .05$).

In the DeepSeek-simulated sample, regional differences reached significance for all variables (Conscientiousness: $F(6, 2936) = 4.22$, $p < .001$, partial $\eta^2 = .01$; Extraversion: $F(6, 2936) = 20.52$, $p < .001$, partial $\eta^2 = .04$; Agreeableness: $F(6, 2936) = 2.26$, $p < .05$, partial $\eta^2 = .005$; Neuroticism: $F(6, 2936) = 15.96$, $p < .001$, partial $\eta^2 = .03$; Openness: $F(6, 2936) = 20.97$, $p < .001$, partial $\eta^2 = .04$; SWB: $F(6, 2936) = 11.51$, $p < .001$, partial $\eta^2 = .02$). Notably, although the overall test for Agreeableness was significant, the effect size was trivial (partial $\eta^2 = .005$) and post-hoc comparisons did not reveal clear pairwise differences. Tukey tests further showed that East China generally scored higher than most other regions on Conscientiousness, Extraversion, (lower) Neuroticism, and Openness. Specifically, East China scored significantly higher in Extraversion than the Northeast, Northwest, and Southwest (mean differences ranging .12–.28, $ps < .05$), and significantly higher in Openness than the Northwest and Northeast (mean differences = .45 and .36, $ps < .001$). For Neuroticism (i.e., lower Neuroticism = greater emotional stability), East China scored significantly better than the Northeast and Northwest (mean differences = .17 and .21, $ps < .05$). For SWB, East China was also significantly higher than Central, Northwest, and Southwest China (mean differences = .24, .29, and .18, $ps < .05$). Meanwhile, the Northeast scored significantly lower than East, North, and South China on both Extraversion (mean differences = .18, .12, .25; $ps < .05$) and SWB ($ps < .05$).

Overall, both the human and simulated samples exhibited significant regional differences in dimensions such as Extraversion, Agreeableness, and SWB. For example, in the simulated sample, East China tended to score higher on Extraversion, Openness, and SWB, whereas the human sample showed particularly elevated SWB in the Northeast (see Figure 3).

Table 2. Means (M) and standard deviations (SD) of Big Five and well-being across seven regions for human and simulated (AI) samples.

| Region | Category | Count | Age | C | E | A | N | O | SWB |
|---|---|---|---|---|---|---|---|---|---|
| NC | Human | 469 | 42.64(13.93) | 3.47(0.65) | 2.99(0.71) | 3.60(0.58) | 2.67(0.71) | 3.25(0.88) | 7.64(2.22) |
|  | AI | 469 | 42.55(12.07) | 3.51(0.37) | 2.31(0.54) | 3.82(0.33) | 3.02(0.42) | 2.82(0.85) | 6.46(0.85) |
| NE | Human | 273 | 46.15(13.31) | 3.46(0.68) | 2.92(0.74) | 3.62(0.58) | 2.60(0.77) | 3.14(0.90) | 7.71(2.13) |
|  | AI | 273 | 44.29(11.91) | 3.49(0.33) | 2.19(0.49) | 3.81(0.30) | 3.12(0.40) | 2.61(0.71) | 6.24(0.68) |
| EC | Human | 700 | 43.62(13.25) | 3.47(0.66) | 2.99(0.72) | 3.48(0.54) | 2.60(0.72) | 3.19(0.81) | 7.49(2.1) |
|  | AI | 700 | 42.29(11.45) | 3.57(0.33) | 2.37(0.52) | 3.85(0.32) | 2.95(0.41) | 2.97(0.75) | 6.52(0.78) |
| CC | Human | 299 | 43.68(13.02) | 3.52(0.63) | 2.98(0.64) | 3.57(0.51) | 2.58(0.76) | 3.17(0.79) | 7.52(2.06) |
|  | AI | 299 | 43.43(13.18) | 3.5(0.37) | 2.28(0.51) | 3.81(0.29) | 3.05(0.39) | 2.71(0.79) | 6.28(0.71) |
| SC | Human | 281 | 40.44(14.06) | 3.46(0.60) | 3.01(0.68) | 3.46(0.50) | 2.60(0.66) | 3.13(0.83) | 6.95(2.14) |
|  | AI | 281 | 43.96(11.50) | 3.52(0.32) | 2.44(0.53) | 3.83(0.3) | 2.96(0.42) | 2.88(0.74) | 6.51(0.82) |
| SW | Human | 467 | 42.17(13.51) | 3.47(0.65) | 3.12(0.66) | 3.49(0.58) | 2.66(0.70) | 3.17(0.82) | 7.14(2.33) |
|  | AI | 467 | 42.24(11.48) | 3.48(0.34) | 2.25(0.54) | 3.86(0.33) | 3.07(0.43) | 2.73(0.71) | 6.34(0.72) |
| NW | Human | 454 | 42.70(12.93) | 3.52(0.66) | 2.92(0.73) | 3.56(0.62) | 2.71(0.76) | 3.29(0.87) | 7.13(2.41) |
|  | AI | 454 | 43.15(11.61) | 3.51(0.36) | 2.09(0.49) | 3.88(0.32) | 3.16(0.39) | 2.52(0.67) | 6.23(0.72) |

Notes: (1) Big Five dimension range 1–5; higher scores indicate stronger trait expression, with higher neuroticism = less emotional stability. (2) Well-being range 1–10; higher = greater well-being. (3) "AI" denotes DeepSeek-generated virtual samples matched on gender/age.

3.3 Relationships between personality and well-being

We conducted multiple linear regressions separately for the human data and the DeepSeek-simulated data to test the predictive effects of the Big Five on SWB. Results are shown in Table 3. In the human sample, the overall model was significant, explaining 3.4% of the variance in SWB ($R^2 = .034$), $F(5, 2937) = 20.58$, $p < .001$. Conscientiousness ($\beta = .23$, $p < .001$), Extraversion ($\beta = .22$, $p < .001$), and Openness ($\beta = .23$, $p < .001$) positively predicted SWB, whereas Neuroticism negatively predicted SWB ($\beta = -.23$, $p < .001$); Agreeableness was not a significant predictor.

In the DeepSeek-simulated sample, the overall model was also significant, explaining 30.0% of the variance in SWB ($R^2 = .300$), $F(5, 2937) = 252.20$, $p < .001$. Openness ($\beta = .28$, $p < .001$) and Agreeableness ($\beta = .13$, $p < .05$) were positive predictors; Neuroticism was a strong negative predictor ($\beta = -.91$, $p < .001$); Extraversion predicted SWB negatively ($\beta = -.28$, $p < .001$); and Conscientiousness was not significant ($p = .836$). These patterns indicate marked structural differences between the AI-simulated and human data in modeling SWB.

To further compare overall structural differences, we conducted principal component analysis (PCA) on the six variables in both samples and plotted radar charts of variable loadings for the first two components (PC1 and

PC2; see Figure 4). The two datasets showed broadly similar structure on PC1, with SWB, Conscientiousness, and Agreeableness being the major loading variables in both. However, on PC2 the loading patterns diverged substantially—especially for Extraversion, Openness, and Agreeableness—with different weightings and even directions between the AI-simulated and human samples, revealing current limitations of LLMs in simulating complex psychological structure.

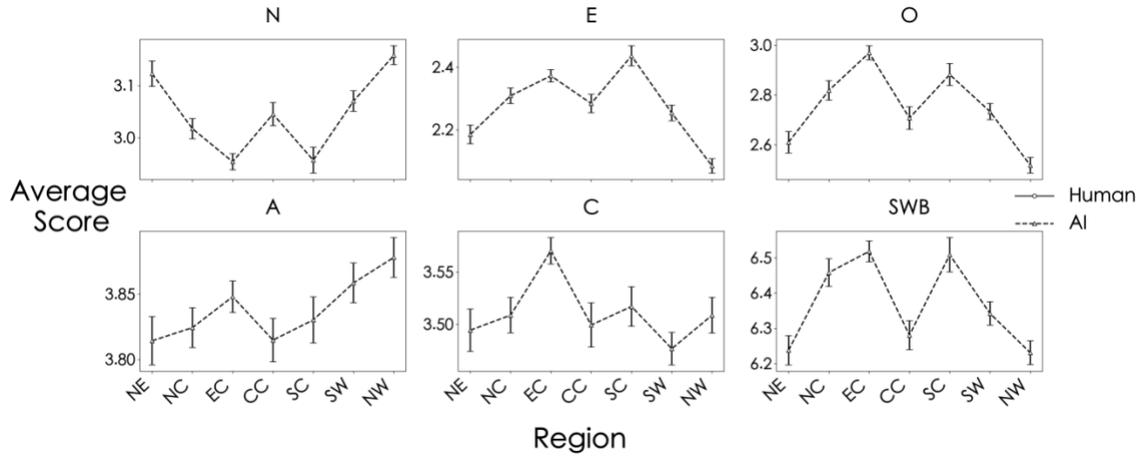

Figure 3. Regional comparison of human vs. simulated data (Big Five and well-being).

Table 3. Multiple regression predicting well-being from Big Five dimensions

| Predictor | β(Human) | SE | t | p | β(AI) | SE | t | p |
|---|---|---|---|---|---|---|---|---|
| Constant | 5.48 | .37 | 14.75 | < .001 | 8.52 | .23 | 36.51 | < .001 |
| C | .23 | .06 | 3.56 | < .001 | .01 | .04 | .21 | .836 |
| E | .22 | .06 | 3.66 | < .001 | -.28 | .04 | -7.80 | < .001 |
| A | .09 | .08 | 1.13 | .258 | .13 | .04 | 3.09 | < .01 |
| O | .23 | .05 | 4.56 | < .001 | .28 | .02 | 12.15 | < .001 |
| N | -.23 | .06 | -4.09 | < .001 | -.91 | .04 | -25.59 | < .001 |

Note: In the table, β denotes the standardized regression coefficient; SE denotes the standard error; t denotes the t statistic for testing the regression coefficient; and p denotes the p-value (two-tailed).

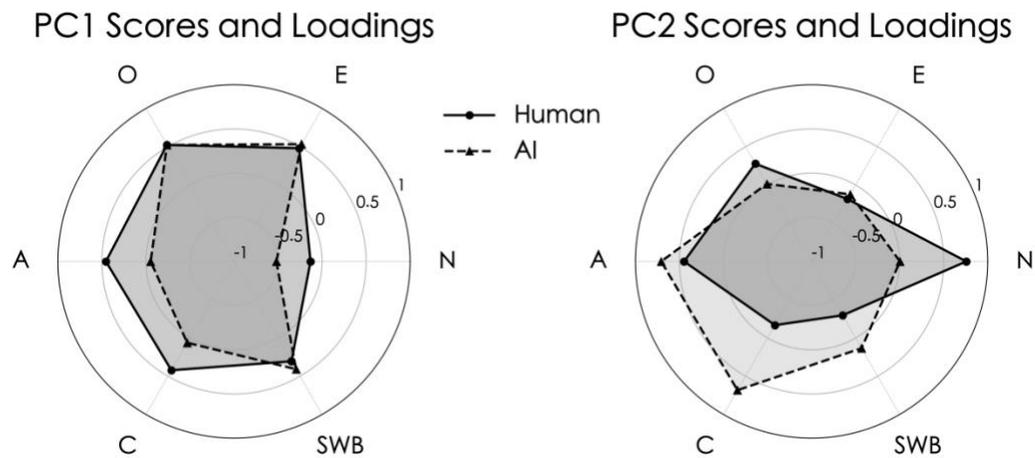

**Figure 4. Comparison of loading structures in principal component analysis (PCA) for the DeepSeek-simulated and Human samples**

Note. The left panel shows the first principal component (PC1), and the right panel shows the second principal component (PC2). The figure displays the standardized loadings of the Big Five dimensions and SWB in the two samples. In the radar charts, the gridlines indicate the magnitude of each variable's weight on the component; larger values indicate greater contribution.

## 4. Discussion

4.1 Summary of findings

We used DeepSeek to simulate regional psychological characteristics. Key findings: (1) Simulated nationwide means/variances of Big Five and well-being were close to real data, indicating the model can reproduce aggregate distributions at the group level. (2) The model broadly reproduced the overall pattern of regional differences, but deviated on specific dimensions (e.g., certain traits and well-being), revealing limitations. (3) Simulated data partially reproduced associations among the Big Five and well-being seen in real human data. Overall, the results support that large language models exhibit good stability and reproducibility on structurally strong, semantically well-defined psychometric dimensions (e.g., the Big Five), consistent with prior findings on LLM-generated virtual participants (De Winter et al., 2024; Jiang et al., 2024). Nevertheless, discrepancies remain for more subjective dimensions such as SWB and for fine-grained details like specific regional rankings.

4.2 Contributions

*4.2.1 Differences between AI simulations and human data*

This study used the large language model DeepSeek to simulate virtual "participants" across China's seven macro-regions in order to assess its ability to represent regional psychological structure. The results indicate that although DeepSeek broadly matches the human sample in demographic composition, there are clear inaccuracies in its simulation of scores on the Big Five dimensions and subjective well-being (SWB). Specifically, significant differences emerged on personality dimensions—especially Extraversion, Agreeableness, Openness, and Neuroticism. For example, Extraversion scores were generally lower in the simulated than in the human sample, with the most pronounced gap in the Northwest (see Figure 1). This suggests that, despite strong text-generation capabilities, LLMs may be constrained by their training corpora when simulating socially oriented traits, making it difficult to faithfully capture human behavior across different social environments.

In addition, the simulated sample tended to score higher than the human sample on Agreeableness and Neuroticism (see Figure 2), which may reflect an over-regularization or exaggeration of affective and social traits in the model's generation process. For SWB, overall scores also differed significantly; in the Northeast, the simulated sample's SWB was notably lower than the human value (see Figure 1), a discrepancy that warrants attention. Human SWB is influenced not only by personality (Anglim et al., 2020; Grant et al., 2009; Zhai et al., 2013) but also by cultural, economic, and social factors (Oishi et al., 2011), and DeepSeek's training data may not sufficiently capture these region-specific influences, limiting its ability to simulate regional SWB. In sum, LLMs can approximate overall trends in psychological characteristics, yet fine-grained discrepancies remain.

Research using LLM-generated virtual participants has yielded preliminary results in personality (De Winter et al., 2024; Jiang et al., 2024), abilities (Mei et al., 2024; Trott et al., 2023), and attitudes (Bisbee et al., 2024), with personality measurement being particularly prominent. Extending this line of work by modeling regional psychological traits with DeepSeek, we find that group-level patterns broadly align with those of human participants—consistent with the individual-level conclusions of De Winter et al. (2024) and Jiang et al. (2024) based on ChatGPT—indicating that LLMs exhibit good stability and reproducibility when simulating personality dimensions that are normatively defined and semantically clear.

*4.2.2 Simulating regional differences*

*Across both the human and DeepSeek-simulated samples, mean levels of subjective well-being (SWB) differed significantly by region (see Figure 3). This indicates that regional variation in SWB among Chinese residents is indeed pronounced, and DeepSeek partially recapitulates this overall trend. However, the fine-grained pattern of differences does not fully align between the simulated and human data. For example, in the human data the Northeast shows significantly higher mean SWB than other regions (see Figure 3), plausibly reflecting stronger community ties and social support networks; yet in the simulated data the Northeast's SWB is clearly underestimated (see Figure 3). Conversely, in the economically developed East China region, the simulated SWB scores are generally inflated (see Figure 3), perhaps because the model over-relies on economic indicators to infer well-being while insufficiently accounting for cultural milieu and social support. Thus, although the model captures the existence of regional disparities in SWB, it does not precisely reproduce the specific ranking of regions.*

*With respect to regional differences in personality traits, analyses of both the human and simulated data reveal certain regional patterns, broadly consistent with prior research on cultural regionalism (蔡华俭等, 2020; 张海钟等, 2012), underscoring the role of cultural background in shaping personality. For instance, our human data show that the Southwest scores significantly higher on Extraversion than other regions (see Figure 3), which may relate to a more open, interaction-rich social atmosphere there. The comparatively higher SWB in the Northeast (see Figure 3) likewise supports the influence of regional culture and social environment on psychological characteristics. Nevertheless, DeepSeek's simulation of these regional personality differences has limitations: the model fails to accurately reproduce some regions' relative strengths or weaknesses. In the simulated data, between-region gaps in Extraversion and Openness are attenuated (see Figure 3), and East China's scores are uniformly elevated across traits (see Figure 3). This aligns with evidence that simulation difficulty varies across personality dimensions—traits such as Extraversion and Openness, which hinge on social interaction and experiential exploration, tend to show lower stability in model outputs (Serapio-García et al., 2023; Sorokovikova et al., 2024; Trott et al., 2023), and cannot be fully emulated by current models. Similarly, the observed biases in simulating regional characteristics reflect structural limitations stemming from LLMs' lack of real experience and interaction within regional cultural and social contexts (Grossmann et al., 2023). In sum, DeepSeek approximates the broad regional patterning of personality traits but remains insufficiently precise in reproducing specific local profiles.*

4.2.3 Predictive Effects of Personality Dimensions on Subjective Well-Being (SWB)

Do the associations between personality and SWB in the DeepSeek-simulated data align with those in real populations? Comparing the predictive effects of the Big Five across the human and simulated samples shows that, in the human sample, Conscientiousness, Extraversion, Openness, and Neuroticism significantly predict SWB (see Table 3). In the simulated sample, Extraversion, Openness, Agreeableness, and Neuroticism significantly predict SWB, whereas Conscientiousness does not (see Table 3). Notably, Neuroticism is negatively associated with SWB (see Table 3) in both datasets, consistent with the robust influence of emotional stability (low Neuroticism) on well-being (陈灿锐等, 2012). However, the DeepSeek data did not reproduce the significant positive effects of Conscientiousness and Extraversion on SWB observed in the human data (Anglim et al., 2020). Specifically, DeepSeek captured the impacts of some traits (e.g., Openness, Neuroticism) more faithfully, but was less accurate for others (see Table 3): the negative effect of Neuroticism on SWB was amplified, the positive effect of Agreeableness was overestimated, and Conscientiousness was not a significant predictor in the simulated sample.

These discrepancies may stem, on the one hand, from the model's reliance during training and response generation on more overt, affect-laden traits, making it difficult to recover the real influence of more implicit, long-term stable traits on SWB. On the other hand, they likely reflect the absence of real interaction and subjective experience in LLMs, which hampers accurate simulation of variables involving subjective feelings—such as SWB and Extraversion (Grossmann et al., 2023). The structural biases we observe accord with recent empirical findings that using LLMs as virtual participants can lead to identity "flattening" and amplified errors in modeling group psychological structure (Wang et al., 2025).

4.3 Applications and Practical Value

This study innovatively introduces large language models into research on regional psychological structure and, without inputting or steering with any CFPS 2018 raw data[4], finds that DeepSeek can reproduce trends in group-level personality and well-being that are consistent with those of human respondents. The work points to promising uses of "virtual participants" to support psychological research and raises methodological considerations for applying this approach.First, discovering research questions. By using DeepSeek to simulate responses from populations in different regions, researchers can anticipate score distributions before fieldwork. This helps identify potential issues in questionnaire design (e.g., culturally unsuitable items) and refine measurement tools, thereby improving validity and reducing the blind spots of on-site surveys.Second, optimizing research design. Prior to large, cross-regional studies, DeepSeek can role-play virtual participants to simulate outcomes under varying conditions. Such pre-experiments allow researchers to probe hypotheses and fine-tune designs—for example, adjusting question wording across contexts and observing model responses to anticipate likely human response patterns. This saves time and cost and offers an initial check of research ideas without adding participant burden.Third, addressing sampling constraints. Empirical studies are often limited by access to samples—insufficient regional cases, or ethically hard-to-obtain variables. DeepSeek can generate complementary virtual data to fill gaps. For instance, when sampling in a given province is difficult, the model can simulate local personality and SWB data to complete the overall analysis. Crucially, such use should be cautious: model outputs must be compared against available human data to ensure reliability; they cannot replace real data and should serve only as auxiliary reference.

---

4. Given the open-ended composition of LLM training corpora, we cannot fully rule out the possibility that the model's outputs were influenced by indirect information related to CFPS 2018 (e.g., scholarly publications or online discussions). Nevertheless, the present study focuses on testing whether the model can simulate real-world distributions of psychological structure without relying on study-specific details. Further tracing of knowledge attribution and the particulars of the training data falls beyond the scope of this paper.

4.4 Limitations and future directions

Overall, as a large language model, DeepSeek shows some potential for simulating the regional distribution of the Big Five and well-being, but also exhibits clear limitations. On the positive side, the DeepSeek simulated sample resembles the real sample in overall demographic composition, and it achieves preliminary success in capturing the broad pattern of regional psychological differences—supporting the feasibility and efficiency of using virtual participants in large-scale psychological research. However, there are several limitations in study design, data sources, simulation strategy, and result interpretation.

First, this study employed a cross-sectional design that generated "virtual participants" with an LLM and compared the simulated outputs to real data. Such a design is essentially correlational; it cannot establish causality and focuses only on regional averages, potentially overlooking within-region individual differences and the dynamic nature of psychological states. Second, the model's pretraining corpora lack sufficiently rich information about regional cultural ecologies and socio-economic contexts, limiting sensitivity to regional psychological differences; at the same time, the real survey data may contain sampling bias or limited representativeness, both of which can affect the reliability of the comparison. Third, DeepSeek and related LLMs have limited capacity for handling complex affective factors and specific cultural contexts. Their simulation of subjective experiences such as well-being is imprecise, and the resulting distributions of psychological traits cannot fully reproduce the richness and diversity of real populations.

Moreover, in interpreting results, partial consistency between simulated and real data does not directly imply that the model has reproduced human psychological mechanisms. Similarities may partly stem from patterns or biases already present in the training corpora; thus, caution is warranted when interpreting findings, and models cannot currently replace real surveys (Dillion et al., 2023; Harding et al., 2024). Finally, an important limitation is the risk that using LLMs to generate virtual participants introduces multi-dimensional stereotyping and structural bias. Because DeepSeek's training data are drawn largely from public internet text—often infused with preconceived regional views about economic development, cultural climate, or social characteristics (Lucy & Bamman, 2021)—the model may inadvertently amplify these stereotypes when simulating regional psychological characteristics, leading to exaggerated deviations for certain regions (Argyle et al., 2023). For example, in this study the Northeast's well-being was markedly underestimated, possibly reflecting reinforced negative narratives in the corpus (e.g., economic downturn, population outflow); conversely, uniformly elevated scores for personality and well-being in East China may reflect overly positive stereotypes in the data. In addition, there is a "visibility" bias in simulating psychological structure: the model more readily captures and magnifies salient, easily verbalized traits (e.g., Neuroticism, Openness), while underperforming on more implicit, long-term stable traits such as Conscientiousness. This economic–cultural–psychological triple bias can introduce systematic error in regional psychological modeling (Wang et al., 2025).

Future research can improve DeepSeek and similar models in several ways. First, enrich and diversify pretraining corpora by incorporating more text about regional cultural ecologies and socio-economic backgrounds (Demszky et al., 2023) to increase sensitivity to regional differences. Second, enhance the simulation of complex emotions and social interaction—for example, by integrating finer-grained affective computing models or long-horizon human affect time-series—to improve the accuracy of modeling well-being and emotional change. Third, conduct cross-cultural comparative studies, integrating samples from multiple countries and regions to test the applicability of LLMs across cultural contexts and to further validate and extend their psychological simulation capabilities. Methodologically, further explore the extensibility of using DeepSeek to simulate group-level psychological data: can its performance on personality and well-being generalize to other psychological processes? Which types of psychological variables resist simulation? These questions merit further investigation.

In sum, this study introduced DeepSeek to simulate regional psychological structure in China, offering preliminary validation of LLMs as "virtual participants" while also revealing current limitations. The findings invite methodological and theoretical reflection: on one hand, LLMs may become innovative tools for reducing the cost of large-sample surveys; on the other, their shortcomings in simulating complex human psychology and cultural context must be squarely acknowledged (Grossmann et al., 2023). With continued enrichment of training data and algorithmic advances, LLMs' ability to simulate regional psychological differences and complex personality characteristics is likely to improve. We look forward to progress in inter-model comparisons, cultural localization, and cross-cultural personality simulation, which could bring methodological transformation to psychological science and provide stronger support for cross-cultural research in personality and social psychology.

## 5. Conclusion

This study innovatively employs the large language model DeepSeek to investigate regional psychological structures. Methodologically, it introduces to psychology a new paradigm for studying regional differences using virtual participants. Substantively, it offers the first validation of how a large real-human sample and an AI-simulated sample perform on social psychological structures.

The findings are as follows: (1) DeepSeek-simulated data show strong group-level modeling capacity for personality traits and well-being, exhibiting overall trends broadly consistent with those of the real sample; (2) at the regional level, the simulated data reproduce the distributional pattern of the Big Five observed in human data, yet deviations remain in specific regional differences for well-being and extraversion, indicating room for improvement in fine-grained modeling; (3) regarding the association between different personality dimensions and well-being, some relational patterns are successfully captured—such as the negative predictive effect of neuroticism—while others diverge, for example the absence of an effect for conscientiousness. In sum, this study not only verifies the potential of large language models for simulating psychological structure, but also provides empirical grounding and theoretical insight for the development of interdisciplinary approaches to psychological measurement.